\documentclass[12pt]{article}
\usepackage{aaspp4}

\newcommand{\gtapprox}{\raisebox{-0.5ex}{$\,\stackrel{>}{\scriptstyle
\sim}\,$}}
\newcommand{\ltapprox}{\raisebox{-0.5ex}{$\,\stackrel{<}{\scriptstyle
\sim}\,$}}
\newcommand{\sigmaT}{\sigma_{\scriptscriptstyle {\rm T}}}
\newcommand{\nciv}{n_{\scriptscriptstyle {\rm CIV}}}
\newcommand{\chiciv}{\chi_{\scriptscriptstyle {\rm CIV}}}
\newcommand{\vj}{v_{\rm jet}}
\newcommand{\vbal}{v_{\scriptscriptstyle {\rm BAL}}}
\newcommand{\vinfty}{v_{\scriptscriptstyle \infty}}
\newcommand{\ro}{r_{\scriptscriptstyle 0}}
\newcommand{\ropc}{r_{{\scriptscriptstyle 0}, \rm pc}}
\newcommand{\roopc}{r_{{\scriptscriptstyle 0}, .1\rm pc}}
\newcommand{\tKH}{t_{\scriptscriptstyle {\rm KH}}}

\received{}
\revised{}
\accepted{}

\lefthead{Kuncic, Z.}
\righthead{BAL~QSOs and the RL/RQ Dichotomy}

\begin{document}

\title{Broad Absorption Line Quasars and the Radio--Loud/Radio--Quiet
Dichotomy}
\author{Zdenka Kuncic\altaffilmark{1}}
\affil{Department of Physics \& Astronomy, University of Victoria,
B.C., V8W 3P6, Canada.}
\altaffiltext{1}{Email: {\tt zdenka@uvastro.phys.uvic.ca}.}

\begin{abstract}
The observation that the extremely broad, blueshifted absorption
troughs which characterize broad absorption line quasars (BALQs)
occur exclusively in radio--quiet quasars (RQQs) suggests that
this class of active galactic nuclei (AGN) may offer important
clues to the radio--loud/radio--quiet (RL/RQ) dichotomy in quasars.
Interestingly, there is also substantial observational evidence for
similar, but lower velocity intrinsic absorption outflows in some
Seyferts and radio--loud quasars (RLQs) as well.
Theoretically, however, it is difficult to interpret this broad range of
mass ejection phenomena in the context of the standard model for BALQs.
Thus, a new model is considered here in which the thermal gas producing
the blueshifted absorption troughs is associated with a poorly--collimated
outflow of weakly radio--emitting plasma -- in essence, a weak jet.
This model provides an appropriate framework not only for assessing
the possible connection between the BAL phenomenon in RQQs and related
intrinsic absorption outflows in stronger radio sources and in less
luminous sources, both of which are known to possess jet--like radio
structure, but also for understanding the RL/RQ dichotomy in light of
recent observations which indicate that at least some RQQs possess central
engines that are capable of producing weak versions of the powerful radio
jets characteristic of RLQs.
In the context of a weak jet model for BALQs, it is shown that
observational constraints on the physical properties of the
radio--emitting plasma are consistent with other theoretical arguments
suggesting that the differences amongst RL and RQ sources can be attributed
to jets with intrinsically different physical properties.
Similarly, theoretical constraints on the physical properties of
absorbing clouds embedded in weak jets are shown to be consistent
with the properties directly inferred from the observed BAL troughs.
Most importantly, however, it is argued that a weak jet model provides
a successful explanation for the anticorrelation between the terminal
velocity of the absorption outflow and the radio power of the quasar.
\end{abstract}


\section{Introduction}

In the framework of the standard accreting black hole paradigm,
unified models have been remarkably successful in explaining the
apparently disparate sub--classes of AGN as simply different facets
of what are otherwise more or less fundamentally identical systems
(\cite{Antonucci93,UrryPad95}).
Despite this success, however, a single, orientation--based scheme
cannot explain the bimodality in the radio luminosity distribution
(\cite{Kellermann89,MillRawSaund93}), so that a true, `grand
unification' scheme for all AGN still remains elusive.
One clue to understanding the RL/RQ dichotomy and unifying AGN is
the possibility that RQQs are capable of producing radio jets,
albeit much weaker, smaller--scaled versions of the powerful,
highly--collimated jets that are characteristic of RLQs.
Some observational evidence supporting this possibility has emerged:
high--resolution imaging of RQQs has revealed that the radio emission
of at least some of these objects originates within a compact, nonthermal
source directly associated with a central engine which appears
qualitatively similar to those in RLQs (\cite{BlundBeas98a,Kukula98});
a correlation between radio and [{\small OIII}]$\lambda 5007$ luminosities,
indicative of the presence of jets, has been measured in not only in
Seyferts (\cite{Whittle85}), but also in some quasars (Miller et al.~1993);
and finally, a significant number of radio--intermediate quasars (RIQs) have
been found, possessing radio emission which is unusually high for RQQs
(but still below that of RLQs) and which has been attributed to weak
beaming (Miller et al. 1993, \cite{FalckSherPat96}), although this still
remains unclear.
But perhaps the most compelling evidence is the recent discovery of
apparent superluminal motion in a RQQ (\cite{BlundBeas98b}).
This is the first direct evidence for a fundamental similarity in
the origin of radio emission in RLQs and RQQs.

Another observational clue to understanding the RL/RQ dichotomy may be
provided by BALQs.
These sources, which comprise $10 - 15 \%$ of optically--selected quasars,
are characterized by (rest--frame) UV spectra with extremely broad (up to
30,000~km.s$^{-1}$) and often deep absorption lines that are blueshifted
with respect to their corresponding emission lines (chiefly resonance lines
due to highly--ionized species such as {\small CIV, SiIV} and {\small NV}).
According to the standard model (\cite{Weymann91} -- and see \cite{Weymann97}
for a recent review), these absorption troughs are formed in a narrow,
quasi--equatorial outflow of numerous, dense ($\sim 10^{6-9}$~cm$^{-3}$)
cloudlets accelerated by line radiation pressure across a region $\sim 1$~pc
in extent.

Although BALs have only been detected in RQQs (e.g. \cite{Weymann91,Stocke92}),
there are two intriguing findings which have yet to be explained: the
statistically significant overabundance of BALQs amongst RIQs
(\cite{Stocke92,FranHoopImp93}); and the recent discovery of `weak' BALs
in a handful of RLQs identified by the {\small FIRST} survey
(\cite{Brotherton98}).
The BALs in these RLQs are `weak' in the sense that the profile widths
of the absorption troughs (several thousand km\,s$^{-1}$) are noticeably
narrower than those typically measured in BALQs and similarly, the
`balnicities'\footnote{The balnicity of an absorption feature is an index
defined by \cite{Weymann91} which depends on the width of the absorption
feature as well as its position with respect to the corresponding emission
line and which thereby measures the likelihood that the feature is a true
BAL rather than due to intervening or associated absorption systems.} are
much lower.
Thus, as pointed out by Weymann (1997), there is an anticorrelation
between the terminal velocity of thermal gas ejected from quasar nuclei
and the radio power of the quasar.
Note, however, that none of the RL~BALQs appear to be powerful radio sources;
all have a ratio, $R^{\ast}$, of radio-to-optical flux (K--corrected)
satisfying $1 \ltapprox \log R^{\ast} \ltapprox 2.5$, while the maximum flux
density measured is no higher than 30~mJy at 20~cm.
Nevertheless, even if these sources are found to more closely resemble RLQs
with `associated absorbers' (\cite{Foltz88}), they can still provide important
clues to the BAL phenomenon since there is evidence that at least some of the
associated absorbers seen in RLQs (mainly steep--spectrum sources) are closely
connected with the nucleus and may perhaps represent the low velocity end of
intrinsic absorption outflows in quasars (e.g. \cite{BarlSarg97,Aldcroft97}).

Similarly, narrow UV absorption features have also been detected in
some Seyfert~1s and these have been interpreted as a low luminosity version
of the BAL phenomenon in quasars.
Moreover, some of these Seyferts (and a few quasars as well) exhibit `warm
absorber' X--ray signatures and there is growing evidence that the outflowing,
absorbing material responsible for the BAL--like features is also responsible
for the X--ray features  (\cite{Crenshaw98,MathWilkElv98,Gallagher99} and
references therein).
It is also interesting to note that some of the nearby AGN exhibiting
nuclear absorption outflows (e.g. NGC~3516, NGC~4151, NGC~5548, Mrk~231)
also exhibit the linear and extended radio structures that are often detected
in Seyferts and that are believed to result from the outflow of radio plasma
along axes determined by the dust torus obscuring the active nucleus
(e.g. \cite{Baum93}).
Indeed, such extended radio structures in Seyferts are generally interpreted
as small--scale, low--power versions of the large--scale, powerful jets and
lobes seen in radio galaxies and quasars (e.g. \cite{UlvWils89}).

Although these observations suggest that nuclear absorption outflows
comprise a dynamically important mass--loss component in AGN that can
span a wide range of parameter space, they are difficult to interpret
in the context of the standard model for the BAL phenomenon in RQQs.
In this paper, the BAL pheonomenon is examined in the context of a
broader model in which nuclear absorption outflows are associated
with poorly--collimated, {\em weak} radio jets.
This model provides a framework not only for examining the connection
between BALs in RQQs and the weaker absorption features in stronger
radio sources (the RL~BALQ candidates) and in lower luminosity counterparts
(Seyferts), but also for testing the hypothesis that all AGN possess
jet--producing central engines and that jets with intrinsically
different physical properties (e.g. radio power, bulk speeds) are at least
partly responsible for the observed RL/RQ dichotomy.
The model is qualitatively outlined in \S~\ref{sec:model} and is then used
to obtain observational and theoretical constraints on the relevant physical
properties of nuclear absorption outflows in \S~\ref{sec:obs} and
\S~\ref{sec:theory}, respectively, and the main results are summarized
in \S~\ref{sec:conc}.

\section{A Weak Jet Model}
\label{sec:model}

In the model constructed here, a weak jet is defined as a poorly--collimated
outflow of radio--emitting plasma moving at a low bulk speed, with a
corresponding bulk Lorentz factor $\Gamma \ltapprox$ a few.
Such an outflow may not necessarily satisfy the traditional criterion for
the formal definition of `jet' (length-to-width ratio $\gtapprox 4$ --
\cite{BridPerl84}), nevertheless, it is useful to apply the jet
description in order to determine the extent to which the RL/RQ dichotomy
can be attributed to differences in jet properties, including the relative
quantities of nonthermal and thermal plasma.
Although the conditions under which jets form still remain poorly
understood, recent theoretical results (\cite{Begelman98}) indicate
that a lack of self--collimation may be due to the absence of a
relatively strong, stabilizing poloidal magnetic field component.
Poorly--collimated jets would lack a Doppler--boosted radio core
and would therefore be observed as much weaker radio sources than
highly--collimated jets.
Thus, jet collimation plays an important role in the observed bimodality
in the radio power distribution. 

While it is physically plausible that all jets contain some thermal
matter in the form of dense clumps (\cite{BegelBlandRees84}), exactly
how much is present, relative to the tenuous, synchrotron--emitting
plasma, remains uncertain.
Recently, Celotti et al. (1998) placed some constraints on the amount
of comoving, thermal gas that could exist in the form of cool, dense
clouds embedded within the powerful, highly--beamed relativistic jets
in RLQs and BL~Lacs.
While it was concluded that such material could only be present in such
energetically insignificant quantities so as to preclude observational
detection, this need not necessarily be the case for thermal gas in
mildly--relativistic and sub--relativistic jets in much less powerful
radio sources.
Indeed, it is argued here that the observational evidence for such jets
is precisely the BAL phenomenon.

As discussed further in \S~\ref{sec:obs}, the distribution of relativistic
particles in such jets can be expected to extend down to thermal energies,
so that the mean Lorentz factor, $\langle \gamma \rangle$, of emitting
electrons is much lower than it is in more powerful jet sources.
This in turn is reasonable to expect if {\it in situ} acceleration of
particles to nonthermal energies (on sub--pc scales) is also less efficient.
Under such conditions, a continuous supply of fresh particles is required
to replenish the `dead' particles that have radiatively cooled and to
thereby maintain a constant radio flux.
As shown by Ghisellini, Haardt \& Svensson (1998), electrons at the
low end of their energy distribution can effectively thermalize via
cyclo--synchrotron self--absorption before escaping the source region
on sub--pc scales.
On these scales, the resulting quasi--thermal electrons can then cool
via inverse Compton scattering; the ratio of the cooling timescale
to the escape timescale is $\sim r_{0.1 \rm pc} v_{0.1c} L_{45}^{-1}$
(where $r=0.1 r_{0.1 \rm pc}$~pc is the source size, $v=0.1v_{0.1c}c$
is the bulk flow speed and $L=10^{45}L_{45}$~erg.s$^{-1}$ is the
luminosity).
The thermal gas can then cool further to temperatures well below the local
Compton temperature ($\sim 10^7$~K) provided the gas density is sufficently
high for bremsstrahlung to become more efficient than inverse Compton
cooling; the required densities are $\gtapprox 10^4 L_{45}r_{0.1 \rm pc}
T_7^{1/2}$~cm$^{-3}$.
As will be shown in \S~\ref{s:physconstraints}, this lower limit
corresponds to the critical density discriminating between the nonthermal
and thermal gas phases in an inhomogeneous jet.
Thus, the local accumulations of condensed, thermal gas which arise as a
result of rapid cooling and poor re--acceleration could be identified as
the progenitors of the absorbing cloudlets which emerge on $\sim$pc scales.
If this is indeed the case, then it provides a natural explantion for
why the BAL phenomenon becomes increasingly rarer in more powerful radio
sources, where presumably {\it in situ} particle acceleration is more
efficient (see e.g. \cite{WeyTurnChrist85} for other possible origins,
including entrainment).

Since velocities measured from the observed blueshifted troughs in BALQs
are typically no more than $\sim 0.2c$, the dense cloudlets could
either be moving slower than the bulk velocity in a mildly--relativistic
jet, or they could be comoving with the radio--emitting plasma in a
sub--relativistic jet.
If such cloudlets comprise a significant kinetic energy flux component
in a mildly--relativistic jet and are moving at sub--jet speeds with a
velocity $\vbal$, then the total energy flux is given by
(e.g. \cite{BridPerl84})
\begin{equation}
\frac{L_{\rm jet}}{r^2  \Omega} \, \simeq \, \left[ \, (\Gamma - 1)
\rho_{\rm jet}c^2 +  \Gamma \frac{B_{\rm jet}^2}{8\pi} \, \right]
\Gamma \beta_{\rm jet}c + \frac{1}{2}\epsilon \rho_{\rm cld} \vbal^3 \; ,
\label{eq:Ljet}
\end{equation}
where $\Omega \simeq 2\pi \phi^2$ is the total solid angle subtended by
jets on either side of the nucleus ($\phi$ is the jet opening angle),
$\Gamma \beta_{\rm jet} c$ is the jet speed, $\rho_{\rm jet}$ and
$B_{\rm jet}$ are the comoving jet mass density (assumed to be dominated
by `cold' protons) and magnetic field, and $\rho_{\rm cld}$ is the mass
density of the clouds, which fill a fraction $\epsilon$ ($\ll 1$) of the
jet volume (the tenuous radio--emitting plasma is assumed to pervade the
bulk of the jet volume).
The relativistic gas pressure is assumed to make a negligible contribution
to the total jet energy flux; the limits on the energy density of synchrotron
emitting electrons calculated in the next section confirm that this is a
valid assumption.
If, on the other hand, the clouds and the radio--emitting plasma (plus
magnetic fields) are {\em comoving} in a {\em sub}--relativistic jet
(i.e. with a velocity $\vj = \vbal \ltapprox 0.2c$), then the total jet
energy flux simplifies to
\begin{equation}
\frac{L_{\rm jet}}{r^2  \Omega} \, \simeq \, \left( \frac{1}{2}
\langle \rho_{\rm jet} \rangle \vj^2 + \frac{B_{\rm jet}^2}{8\pi}
\right) \vj
\label{eq:Ljet_sub}
\end{equation}
where $\langle \rho_{\rm jet} \rangle \simeq \rho_{\rm jet} + \epsilon
\rho_{\rm cld}$ is now the {\em average} comoving mass density of the jet.

\section{Observational Constraints}
\label{sec:obs}

\subsection{Covering Factor}

The widely adopted, standard model for BALQs is chiefly founded upon
the observational study by Weymann~et~al.~(1991), who found no statistically
significant differences between the spectral properties of BALQs and
non-BALQs in a sub--sample taken from the LBQS, thus indicating that
BALQs do not form an intrinsically different class of objects from non-BALQs.
When combined with the constraint from scattering models that the BALR cannot
completely occult the continuum source (\cite{Junk83} -- see also
\cite{HamKorMor93}), this result led
to the suggestion that {\em all} RQQs possess a BALR with a global covering
factor that can be identified with the incidence rate of BALQs amongst an
optically--selected sample, typically $\sim 0.1 - 0.15$ (although the
`true' incidence rate could be as high as $30\%$ if attenuation is taken
into account -- see e.g. \cite{SchmidtHines99} and references therein).
It was then further suggested that a physically plausible distribution
for absorbing cloudlets would be a quasi--equatorial geometry, possibly
skimming the edge of an obscuring torus, which would provide a natural
source of material for the cloudlets.

Note that a weak jet model may not necessarily be compatible with this
covering factor interpretation of the BAL incidence rate amongst
optically--bright quasars.
A poorly--collimated jet could, in principle, freely expand to fill the 
biconical regions interior to the dusty torus, so that the half--opening
angle of the `jet' could be as wide as $60^\circ$.
Since optically--bright quasars are also those seen along a direct line
of sight to the continuum source (i.e. within these `ionization cones'),
then the low incidence rate of BALs amongst these quasars may not
necessarily be consistent with the jet covering factor; it then becomes
necessary to consider the BAL phenomenon as an evolutionary, mass--loss
phase (e.g. \cite{Miller97}).
Note that the original Weymann~et~al.~(1991) results do not rule out a
duty cycle effect for the BAL phenomenon (\cite{Weymann97}) and indeed,
there is some observational evidence to support the idea
(e.g. \cite{BrigTurnWolf84,BorPearOke85,VoitWeyKor93})
that BALQs may be transition objects between RQ and RL quasar phases that
have undergone a close interaction and/or merger event which has triggered
the expulsion of excess mass and angular momentum.
One particularly impressive example is the recent adaptive optics image of
the BALQ PG~1700+518 ($z=0.29$), which clearly reveals a discrete companion
galaxy that appears to be merging with the quasar (\cite{StockCanalClose98}).
Similarly, other BALQs which are sufficiently nearby ($z<0.5$) to show
clear signs of having undergone a recent interaction or merger event
include Q~0205+024, IRAS~0759+6508, Q~1402+436 and Q~2141+175, and while
several more show less discernable signs (e.g. PG~0026+129, PG~0043+039,
Q~0318-196, PG~1426+015, PG~2233+143) their immediate environs are strongly
suggestive of interactions taking place.
Finally, tidal tails and nearby companions associated with some low-$z$
quasars have been detected by HST imaging (e.g. \cite{Bahcall97}), which
has also revealed that the host galaxies of at least some bright RQQs, like
those of RLQs and radio galaxies, are massive ellipticals, which are believed
to have been formed from mergers (\cite{McLure98}).

\subsection{Orientation and Geometry}

The strongest evidence for a quasi--equatorial geometry for the BALR has
come from polarization measurements, which have revealed that, on average,
BALQs tend to have higher levels of optical polarization than than non-BALQs
and which, when interpreted in terms of orientation alone, suggest that the
BALR is being intercepted along highly--inclined lines of sight
(\cite{HutsLamRem98,SchmidtHines99} and references therein).
These data have also revealed that the highest levels of polarization
($> 1\%$) are measured in low--ionization BALQs (lo-BALQs; those with
absorption lines due to low--ionization species, such as {\small Mg~II}
and {\small Al~III}, in addition to the usual high--ionization BAL troughs).
Since the objects in this sub--class of BALQs also show evidence of strong
dust reddening (\cite{SprayFoltz92}), the polarization data strongly
favour models in which the viewing angle is close to the obscuring dust
torus.
However, this is strictly only the case for lo-BALQs; polarization
studies have found no statistically significant differences in the optical
polarization between high--ionization BALQs (hi-BALQs) and non-BALQs
(Hutsem\'{e}kers et al.~1998) and therefore, they offer no helpful clues
to the orientation of hi-BALQs, which in fact make up the majority of BALQs.

Another way of determining the orientation and geometry of the BALR is to
search for radio axes.
Unfortunately, the large distances and low radio fluxes of RQQs have made
it difficult in practice  to resolve radio images of BALQs.
Indeed, prior to the recent {\small FIRST} survey, only two BALQs had been
mapped with sufficient spatial resolution with the VLA: PG~1700+518, which
exhibits double compact radio structure down to $0.15''$ at 15~GHz
(\cite{HutchNeffGow92,Kellermann94,Kukula98}); and the Cloverleaf,
H~1413+1143, which exhibits compact radio counterparts to all four of the
optical images produced by gravitational lensing, as well as an additional,
strongly amplified radio source that appears to be associated with the quasar
itself (possibly an ejected radio component; \cite{Kayser90}).
Similarly, the newly discovered BALQ APM~08279+5255 (\cite{Irwin98}) exhibits
double compact radio structure down to $0.28''$ at 3.5~cm (G.F.~Lewis, pvt.
com.).
Even the recent {\small FIRST} survey, which has detected and mapped,
with follow--up, high resolution (A array) VLA imaging, about 20 BALQs,
has failed to detect any elongated or extended structure that could be
identified as radio axes; all of the sources appear point--like down to
a $0.2''$ resolution level (B.~Becker, pvt. com.).
Although it may be possible to interpret these radio sources as weak,
unresolved jets, it would be desirable to obtain higher quality
radio data.

In the meantime, it is interesting to make a comparison with the low
luminosity, low velocity counterparts to BALs found in Seyfert~1s which
are sufficiently nearby to resolve linear radio structures on
sub--kiloparsec and sometimes parsec scales.
For example, NGC~3516, NGC~4151 and NGC~5548, which are classified as
Seyfert~1.5s, all exhibit elongated radio structure with subcomponents
and with an unresolved core centred on the the optical nucleus
(e.g. \cite{Baum93}).
On the other hand, in other nearby AGN which exhibit BAL--like features
(e.g. NGC~3783, NGC~509, NGC~7469), no radio axes are detected, only
nuclear point sources.
This is typically the case for objects which are classified as
Seyfert~1.0-1.2 and which are therefore believed to be viewed at low
inclinations, so it is unclear whether they actually possess linear
radio structure that cannot be seen because of a lack of projection, or
whether their radio sources are intrinsically different from those in
other Seyferts, which seems less likely to be the case.

There are also other observational clues to suggest that not all BALQs
are being viewed at large inclination angles.
For example, dust models for the optical to submillimeter spectral energy
distributions of H~1413+117 and of APM~08279+5255 (both of which are
{\small IRAS} sources) are consistent with a dusty torus being viewed
face--on, with a direct, unobscured line of sight to the optical continuum
source (\cite{Barvainis93,Lewis98}).
Similarly, the lack of reddening in other hi-BALQs (e.g. \cite{Weymann91})
suggests that they too are being viewed at latitudes sufficiently high
to avoid dust contamination from the putative torus.
Furthermore, the remarkable similarity between the emission line
equivalent widths of BALQs and non--BALQs is surprising, given that
projection effects are expected to produce measureable differences
if they are indeed seen from different viewing angles
(P.~Francis, pvt. com.) 
The orientation interpretation of spectropolarimetric data is also unclear;
resonance line scattering by ions in an equatorial geometry is predicted to
produce additional, redistributed polarized flux in the red wings of the
emission lines (\cite{LeeBland97}), but in some cases, a {\em deficit} of
polarized flux redward of the permitted emission lines is detected
(see e.g. Fig.~4 in \cite{Weymann91} and \cite{Ogle97}).
Note that a weak jet, with the properties outlined in \S~\ref{sec:model},
would produce negligible optical flux and therefore, a negligible
contribution to the continuum polarization.

\subsection{Limits from Flux Density Measurements}
\label{sec:fluxdens}

The standard synchrotron formulae for a homogeneous source region can be
used to place some limits on the physical properties of the radio--emitting
plasma and magnetic fields that are capable of producing the observed radio
flux densities of BALQs, which are typically  $\sim$~a few~mJy.
The emitting electrons are assumed to have the usual nonthermal energy
distribution: $n_{\gamma} \propto \gamma^{-p}$, with a total electron number
density $n_{\rm e} = \int {\rm d}\gamma \, n_{\gamma}$, where $\gamma$ is
the electron Lorentz factor, with $\gamma_{\rm min} \ll \gamma \ll
\gamma_{\rm max}$, and where $p$ is the particle spectral index.
For optically--thin synchrotron emission, the spectral index is given by
$\alpha = (p-1)/2$ and the observed flux density can be related to the
other observable parameters, the angular diameter of the source,
$\theta_{\rm d}$, and the luminosity distance, $D$, according to
(see \cite{Marscher87})
\begin{equation}
S_{\nu} \simeq (3 \times 10^4) \gamma_{\rm min}^2 n_{\rm e}
B^2 \nu_{\rm GHz}^{-1} \theta_{\rm mas}^3 D_{\rm Gpc}
(1 + z)^{-2} \;\; \mbox{mJy ,}
\label{eq:Sobs}
\end{equation}
where $B$ is the magnetic field and $\alpha = 1.0$ has been used, since this
is a typical value obtained from observed BALQ radio spectra for which
spectral indices could be measured (\cite{BarvLons97}).
To take into account the possibility that the observed radio flux has been
boosted as a result of beaming (i.e. in RLQs), this expression needs to be
further multiplied by a factor $\delta^4$, where $\delta = [ \Gamma
(1 - \beta \cos \varphi) ]^{-1}$ is the Doppler factor corresponding to a bulk
velocity $\beta c$ with a bulk Lorentz factor $\Gamma$ and with a direction
$\varphi$ with respect to the observer.

To obtain independent constraints on the unknown source parameters $n_{\rm e}$
and $B$, the optically--thin synchrotron spectrum can be extrapolated down to
the frequency $\nu_{\rm m}$ where the observed flux density reaches a maximum
at a value $S_{\rm m}$ and where it can be assumed that the optical depth to
synchrotron self--absorption is approximately unity (see \cite{Marscher87}).
This then gives the following relations:
\begin{equation}
B \simeq 40 \, \left( \frac{\nu_{\rm m}}{\rm GHz} \right)^5
\left( \frac{S_{\rm m}}{\rm mJy} \right)^{-2}
\theta_{\rm mas}^4 \, (1 + z )^{-1} \delta \;\; {\rm G} \;,
\label{eq:Bm}
\end{equation}
which is virtually independent of $\alpha$, and for $\alpha = 1.0$:
\begin{equation}
n_{\rm e} \simeq ( 4\times 10^{-7} ) \gamma_{\rm min}^{-2}
\left( \frac{\nu_{\rm m}}{\rm GHz} \right)^{-9}
\left( \frac{S_{\rm m}}{\rm mJy} \right)^5 \theta_{\rm mas}^{-11} 
D_{\rm Gpc}^{-1} (1+z)^8 \delta^{-6} \;\; {\rm cm}^{-3}
\label{eq:nm}
\end{equation}
Although these relations are strongly dependent on the observable parameters,
they can be somewhat useful when comparing the extremely contrasting
properties between the compact radio cores of RLQs and the much weaker
radio sources in RQQs (including BALQs).
In particular, these relations imply a distinct difference between the
ratio of energy densities in magnetic field, $u_{\rm B}$, to relativistic
electrons, $u_{\rm e}$, for quasars with contrasting radio properties.
Eqn.~(\ref{eq:Bm}) implies a magnetic energy density
$u_{\rm B} = B^2 / 8\pi \sim 50 \, \nu_{\rm GHz}^{10} S_{\rm mJy}^{-4}
\theta_{\rm mas}^8 (1+z)^{-2} \delta^2$~erg.cm$^{-3}$, while eqn.~(\ref{eq:nm})
implies an electron energy density (for $\alpha =1.0$)
$u_{\rm e} = 2\gamma_{\rm min}^2 n_{\rm e} m_{\rm e}c^2 \sim 10^{-12} \,
\nu_{\rm GHz}^{-9}S_{\rm mJy}^5 \theta_{\rm mas}^{-11} D_{\rm Gpc}^{-1}
(1+z)^8 \delta^{-6}$~erg.cm$^{-3}$.
Interestingly, the ratio $u_{\rm B} / u_{\rm e}$ for RQQs is larger by
many orders of magnitude than the same ratio for RLQs (assuming the
same observing frequency and the same redshift), even if a conservative
flux density (say, 100~mJy) and a high Doppler factor ($\delta \simeq 10$)
are used for the RL source.
Also, the condition $u_{\rm B} / u_{\rm e} \gg 1$ is always obtained
for BALQs, even in the case of the highest flux density level measured
so far (for {\small FIRST}~1556+3517; one of the RL BALQ candidates),
30~mJy at 1.4~GHz, with $z=1.48$ (\cite{Brotherton98}), which gives a
lower limit of
$u_{\rm B} / u_{\rm e} \gg 0.1 (\nu_{1.4} \theta_{\rm mas})^{19} D_{\rm Gpc}$,
and which, taking into account the strong dependence on $\theta_{\rm mas}
(\gtapprox 1)$, always exceeds unity by an appreciable amount.

The ratio $u_{\rm B} / u_{\rm e}$ may have important implications for the
nature of the radio emitting source regions in quasars, especially in the
framework of jet models.
Falcke \& Biermann (1995), for instance, suggest that there exists a `family'
of jet models, the members of which are distinguished by differences in the
equipartition conditions involving the energy densities in the magnetic field,
relativistic electrons and protons as well as thermal electrons and protons
and also differences in the total energy budget of the jet--disk system
as a whole.
According to their hypothesis, jets with $u_{\rm B} \gg u_{\rm e}$ are
predicted to be radio quiet if the relativistic electron distribution
begins at $\gamma_{\rm min} \simeq 1$ and if $u_{\rm B}$ is below its
equipartition value with respect to the bulk kinetic energy.
They also further argue that jets with $u_{\rm B} \gg u_{\rm e}$ and
$\gamma_{\rm min} \simeq 1$ can be radio {\em weak} (but not radio
{\em quiet}) if $u_{\rm B}$ is in equipartition with the bulk kinetic energy,
thus offering a plausible theoretical discrimination between RQ and RI sources.
Note that while the idea (e.g. \cite{FalckSherPat96}) that RIQs are
Doppler--boosted RQQs may seem appealing in the framework of a weak jet
model, there is very little observational evidence for beaming in non-RLQs.

\section{Theoretical Constraints}
\label{sec:theory}

Although various pressure--driven wind models have been proposed for BALs
(see \cite{deKool97} for a summary), the only direct observational clues to
the nature of the driving force are line--locking features (\cite{Turnshek88})
and `ghost of Ly~$\alpha$' features (\cite{AravBegel94}).
There are, however, theoretical arguments to suggest that while line
radiation pressure clearly plays an important dynamical role in the BAL
phenomenon, it may not necessarily be the only acceleration mechanism.
For instance, absorbing clouds will experience large forces when
they move relative to an accelerating, confining medium and thus, they
will be unavoidably dragged along by the dynamic pressure of the
external fluid (Weymann et al.~1985).
Indeed, Arav, Li \& Begelman (1994) have shown that BAL clouds comoving
with the ambient medium produce profiles that more closely resemble those
observed than do the profiles produced by line acceleration alone when
the clouds are decoupled from the ambient medium.
They also find that to produce a significant contribution to the
overall acceleration from line pressure relative to ram pressure when the
clouds are comoving, the starting radius is too close to the inferred
radius of the broad emission line region, i.e. $\sim 0.1$~pc.

Another related problem is the cloud confinement mechanism.
The temperatures required for pressure confinement by a thermal wind
(e.g. \cite{Stocke92}) are difficult to achieve on $\sim$~pc scales,
while a wind driven by cosmic rays (e.g. \cite{BegeldeKoolSik91})
also cannot provide the necessary pressure for confinement of BAL clouds.
The confinement problem disappears if, instead of clouds, the BALs are
produced by a quasi--continuous, high column density wind
(e.g. \cite{Murray95}).
Although such a model is made more appealing by being able to account for
the common UV/X--ray (BALs/warm absorber) absorption features that have
been detected in some sources (\cite{Crenshaw98,Gallagher99} and references
therein), it  requires ionization parameters several orders of magnitude in
excess of the values inferred from the range of ionization states in the
observed BAL troughs and this also makes it difficult to account for lo-BALs.

Whether BAL clouds can be accelerated and confined by a weak jet and whether
the physical properties of such clouds are consistent with those deduced from
observations now remains to be determined. 

\subsection{Dynamical Considerations}

Consider a blob of gas immersed in an outflowing medium.
This blob, irrespective of its formation history, will quickly come into
pressure equilibrium with its surroundings and in doing so, will be
accelerated by the dynamic pressure of the outflow, expanding as it moves
downstream.
For a jet of speed $\Gamma \beta_{\rm jet}c$, the ram pressure exerted
on a cloud of scalelength $r_{\rm cld}$ satisfies
\begin{equation}
\rho_{\rm cld} v_{\rm cld} \frac{\partial v_{\rm cld}}{\partial r} =
\frac{\rho_{\rm jet} \Gamma^2 (\beta_{\rm jet}c - v_{\rm cld})^2}
{r_{\rm cld}} \; ,
\label{eq:motion_rel}
\end{equation}
where $v_{\rm cld}$ is the cloud velocity.
However, the momentum flux, $\rho_{\rm jet} \Gamma^2 \beta_{\rm jet}^2 c^2$,
of a sub--relativistic jet is higher than that of a relativistic jet with
the same energy flux, i.e. the ratio of momentum-to-energy flux,
$(\Gamma / \Gamma - 1) \beta_{\rm jet} / c$, is higher by a factor $2c/\vj$
and therefore, the dynamic pressure of a sub--relativistic jet (of speed
$\vj$) provides a more efficient acceleration mechanism than that of a
relativistic jet.
Indeed, ram pressure acceleration in a relativistic jet is no more
efficient than acceleration by line radiation pressure (the favoured
mechanism for BALs) for the same power in kinetic energy flux and photon flux.
In a {\em sub}--relativistic jet, on the other hand, the ratio of
$a_{\rm ram}$ to $a_{\rm rad}$ is $\sim c/v_{\rm jet}$.

The higher efficiency of ram pressure acceleration in a sub--relativistic
jet compared to that in a relativistic jet led \cite{BlandKon79} to predict
that dense blobs embedded in a sub--relativistic jet would naturally give
rise to absorption troughs blueshifted with respect to their corresponding
emission lines.
This also immediately suggests that a jet model offers a natural explanation
for why nuclear absorption outflows, when present in RLQs, are never as
strong as those which characterize bonafide BALQs.

\subsubsection{The $\vinfty$---Radio-Loudness Anticorrelation}
\label{sec:anticorr}

The distinct anticorrelation between the observed terminal velocity,
$\vinfty$, of material ejected from a quasar nucleus and the radio power
of the quasar, as pointed out by Weymann~(1997) following the {\small
FIRST} discovery of BALs in RLQs (\cite{Brotherton98}), is clearly a key
observational property which therefore provides a critical test for the
dynamical aspects of any theoretical model.
It is clearly difficult to interpret this observation in the context of
models in which the momentum of the outflowing gas entirely derives from
the radiation field.
On the other hand, an outflow driven at least partially by the momentum
flux of a radio--emitting jet, whose radio flux is some fraction of the
total energy flux, clearly warrants a more quantitative investigation.

Consider a total column density $N_{\rm cld}$ of absorbing clouds accelerated
along a line of sight by the dynamic pressure of a jet.
These clouds will attain a terminal velocity according to
(c.f. eqn.~\ref{eq:motion_rel})
\begin{equation}
\frac{\vinfty^2}{c^2} \ltapprox \frac{\Gamma \beta_{\rm jet}}{\Gamma - 1}
\frac{2 L_{\rm jet}}{\ro \Omega N_{\rm cld} m_{\rm p}c^3}  \;\; ,
\label{eq:terminalv}
\end{equation}
where $\ro \! = \! \ropc \!$ pc is the radius at which the acceleration
commences and where eqn.~(\ref{eq:Ljet}) has been used.
Thus, a sub--relativistic jet can accelerate a total column density of
$10^{22}N_{22}$~cm$^{-2}$ clouds to comoving velocities
\begin{equation}
\vinfty \, \ltapprox \, 0.1c \, \, L_{46}^{1/3} \, (\ropc N_{22})^{-1/3}
\left( \frac{\Omega}{4\pi} \right)^{-1/3}
\label{eq:v_subrel}
\end{equation} which is consistent with the maximum velocities
measured directly from the blueshifted absorption troughs in BALQs (e.g.
\cite{Weymann91}).
In the case of a mildly--relativistic jet, with, say $v_{\rm jet} \!
= \! 0.5c$ (corresponding to $\Gamma = 1.15$), eqn.~(\ref{eq:terminalv})
implies a terminal velocity much less than the bulk jet speed, with
\begin{equation}
\vinfty \, \ltapprox \, 0.07c \, L_{\rm jet, 46}^{1/2} (\ropc N_{22}
v_{\rm jet,0.5})^{-1/2} \left( \frac{\Omega}{4\pi} \right)^{-1/2}
\label{eq:v_mildrel}
\end{equation}
Note that in the limit of relativistic jet speeds ($\Gamma \gtapprox$ a
few), any dense clouds embedded in the flow will always be accelerated
to the bulk velocity, unless the jet is `free', with an opening angle
$\phi \gg \Gamma^{-1}$ (Begelman~et al.~1984), in which case
eqn.~(\ref{eq:terminalv}) implies
$\vinfty \ll 0.1c L_{46}^{1/2} \Gamma_3 (\ropc N_{22})^{-1/2}$.
Thus, a jet model offers a viable explanation for why the outflow velocities
associated with the much narrower, blueshifted absorption lines in RLQs are
never as high as the velocities associated with genuine BALs in RQQs.

It is also of interest to perform these calculations with lower energy
fluxes to test the applicability of a weak jet model to the UV absorption
features detected in some Seyferts (e.g. \cite{Crenshaw98}).
Using appropriate scaled--down values for $L_{\rm jet}$ and $\ro$ of, say
$10^{43}$erg.s$^{-1}$ and $0.1$~pc, respectively, eqn.~(\ref{eq:v_subrel})
implies comoving velocities $\ltapprox 0.02c L_{43}^{1/3} (\roopc N_{22})
^{-1/3} (\Omega / 4\pi )^{-1/3}$.
This is consistent with observations, which indicate that the absorption
features in Seyfert spectra are never as broad as those seen in their more
luminous counterparts, with terminal velocities of $\ltapprox 0.015c$
typically being measured, compared to $\ltapprox 0.2c$ for the BALQs.

Thus, a jet model for BALQs can not only explain the observed
$\vinfty$--radio-loudness anticorrelation in quasars, but can also explain
the lower $\vinfty$ values measured from weaker absorption features in
less powerful sources.

\subsubsection{Kevin--Helmholtz Instability}

Small--scale clouds moving with a relative velocity with respect to the bulk
velocity of the surrounding plasma are susceptible to the Kevin--Helmholtz
instability, which can shred the clouds into smaller and smaller entities.
For clouds embedded in a magnetized medium, moving with a relative velocity
$\Delta v$, the fastest growth timescale corresponding to the most disruptive
modes is (e.g. \cite{Celotti98}; see also Begelman et~al. 1991)
\begin{equation}
\tKH \simeq \frac{r_{\rm cld}}{\Delta v}
\left( \frac{\rho_{\rm cld}}{\rho_{\rm jet}} \right)^{1/2}
\label{eq:t_KH}
\end{equation}
If the clouds are confined by the dynamic pressure of the jet,
i.e. $\rho_{\rm cld}c_{\rm s}^2 \simeq \rho_{\rm jet}\vj^2$, then
$\tKH \simeq (\vj / \Delta v) t_{\rm sc} \gtapprox t_{\rm sc}$, where
$t_{\rm sc} = r_{\rm cld} / c_{\rm s}$ is the sound--crossing timescale
across the clouds, corresponding to an internal sound speed
$c_{\rm s} = \sqrt{2kT_{\rm cld}/m_{\rm p}}$.
In other words, the instability is sufficiently rapid
to restrict the confinement of clouds to timescales as short as $t_{\rm sc}$.
Since the acceleration timescale is much longer than $\tKH$, this means
that clouds must be continuously regenerated or injected along the outflow.

In the nonlinear regime, the Kevin--Helmholtz instability causes a rapid
cascade of cloud fragmentation.
While this does not directly destroy the clouds, it makes them increasingly
more prone to microphysical diffusion processes, which can assimilate
the clouds into the ambient medium, thereby effectively causing their
evaporation.
This is examined in \S~\ref{s:evaporation} below (see also Weymann et~al.
1985 for a discussion).

\subsection{Physical Constraints}
\label{s:physconstraints}

In the following, it is assumed that the BAL cloudlets are comoving with
the bulk flow of a sub--relativistic jet, since the arguments presented
in Section~\ref{sec:anticorr} indicate that this may be an appropriate model
for bonafide BALQs.
From eqn.~(\ref{eq:Ljet_sub}), the total power in a sub--relativistic
jet can be written as
\begin{equation}
L_{\rm jet} \gtapprox r^2 \Omega \frac{1}{2} \langle \rho_{\rm jet} \rangle
\vj^3 \;\; ,
\label{eq:Ljet_non}
\end{equation}
from which an upper limit on the mean jet density can be obtained:
\begin{equation}
\frac{ \langle \rho_{\rm jet} \rangle}{m_{\rm p}} \ltapprox (4 \times 10^3)
L_{46} \, r_{\rm pc}^{-2} \left( \frac{\vj}{0.1c} \right)^{-3}
\left( \frac{\Omega}{4\pi} \right)^{-1} \;\; {\rm cm}^{-3}
\label{eq:njet}
\end{equation}

\subsubsection{Confinement}
\label{s:confinement}

A crucial issue which needs to be addressed by any physical model for
BALQs is the confinement mechanism.
If BAL clouds are accelerated by the dynamic pressure of a weak jet,
then ram pressure provides a natural confinement mechanism.
This corresponds to the equipartition condition $\rho_{\rm cld}c_{\rm s}^2
\simeq \rho_{\rm jet}\vj^2$.
Using eqn.~(\ref{eq:Ljet_non}), this implies a characteristic cloud density
\begin{equation}
n_{\rm cld} \ltapprox \frac{2L_{\rm jet}}{r^2 \Omega \vj kT_{\rm cld}}
\simeq 10^{10} L_{\rm jet,46}r_{\rm pc}^{-2}
\left( \frac{\vj}{0.1c} \right)^{-1} \left( \frac{\Omega}{4\pi} \right)^{-1}
\left( \frac{T_{\rm cld}}{3 \times 10^4~{\rm K}} \right)^{-1} \;\;
{\rm cm}^{-3} \; ,
\label{eq:ncld}
\end{equation}
which is consistent with the upper limits deduced from the observed
ionization species and from photoionization models
(see e.g. \cite{Turnshek88}).

It is also possible that comoving magnetic fields provide pressure support
to dense cloudlets.
The typical field strength of a comoving magnetic field is
\begin{equation}
B_{\rm jet} \simeq L_{\rm jet , 46}^{1/2} \, r_{\rm pc}^{-1}
\left( \frac{\vj}{0.1c}\right)^{-1/2}
\left( \frac{\Omega}{4\pi}\right)^{-1/2} \;\; {\rm G}
\label{eq:Bjet}
\end{equation}
which satisfies the equipartition condition
$B_{\rm jet}^2 / 8\pi \simeq n_{\rm cld} kT_{\rm cld} \simeq
L_{\rm jet} / r^2 \Omega \vj \simeq \frac{1}{2}\rho_{\rm jet}\vj^2$.
According to Falcke \& Biermann (1995), jets in which the magnetic field
is below equipartition with the bulk kinetic energy are likely to be
radio--quiet sources, so this could be a distinguishing property between
bonafide BALQs and the RL BALQ candidates.
Although magnetic fields need not play an important dynamical role in a
weak jet model for BAL outflows, even small field strengths can be of
crucial importance to maintaining a two--phase fluid by suppressing
transverse diffusion of relativistic particles (whose motion is confined
to a Larmor radius about the field lines) into cool, dense BAL clouds.
However, longitudinal diffusion can still be important and therefore
needs to be examined.

\subsubsection{Evaporation}
\label{s:evaporation}

A serious threat to the survival of BAL clouds embedded in a jet is
evaporation into the ambient plasma as a result of diffusion and Coulomb
heating by the external fast particles which fill the bulk of the jet.
The volume heating rate due to Coulomb collisions between thermal,
nonrelativistic ($kT_{\rm e} \ll m_{\rm e}c^2$) electrons and nonthermal,
relativistic electrons is given by (e.g. \cite{Gould72}; see also
\cite{Jackson75})
\begin{equation}
H^{\rm Coul} \simeq \frac{3}{2} \chi_{\rm e} n_{\rm cld}n_{\gamma}\sigmaT
m_{\rm e}c^3 \, {\cal B} (\gamma) \;\; ,
\label{eq:Coulheat}
\end{equation}
where $\chi_{\rm e}$ is ratio of the number density of thermal electrons
(either free or harmonically bound to ions) in the cloud gas to the total
cloud density and ${\cal B}( \gamma )$ is a parameter which only weakly
depends on $\gamma$ and which is related to the logarithmic Gaunt factor,
determined from the maximum and minimum impact parameters.
For collisions with free electrons, which determine the overall heating of
the cloud gas, ${\cal B}(\gamma )\simeq \ln ( \sqrt{\gamma - 1} \,
m_{\rm e}c^2 / \hbar \omega_{\rm p} )$, where $\omega_{\rm p}$ is the
(thermal) electron plasma frequency.
Eqn. (\ref{eq:Coulheat}) corresponds to a collision timescale
\begin{equation}
t_{\rm coll} \simeq \frac{2 \, \gamma}{3n_{\rm cld} \sigmaT c \cal{B}}
< 10^5 n_{\rm cld,8}^{-1} \, \frac{\gamma}{\cal B} \; {\rm s} \;\; ,
\label{eq:tcoll}
\end{equation}
where $n_{\rm cld,8} = n_{\rm cld} / 10^8$cm$^{-3}$.
This must be appreciably longer than the sound--crossing timescale,
$t_{\rm sc}$, if pressure confinement of individual clouds is to be
sustained in spite of the collisions, implying cloud sizes
\begin{equation}
r_{\rm cld} \ll (5 \times 10^{11}) \, n_{\rm cld,8}^{-1}
\left( \frac{T_{\rm cld}}{3 \times 10^4{\rm K}} \right)^{1/2}
\frac{\gamma}{\cal{B}} \;\; {\rm cm}
\end{equation}
This is smaller than the mean--free--path between the collisions,
$\lambda_{\rm mfp} = t_{\rm coll}\beta c$ (where $\beta c$ is the
velocity of the relativistic electrons) by a factor $\sim c/c_{\rm s}
\sim 10^4$, which means that the relativistic electrons would have to
travel through as many clouds before a Coulomb encounter occurs and
diffusive effects become important.
In other words, the particle energy in the ambient jet plasma is simply
{\em advected} through the clouds, rather than transferred diffusively,
as a result of direct encounters between the thermal and nonthermal
electrons.
Furthermore, this will only strictly be true if the magnetic field lines
in the jet penetrate the clouds with little distortion.
If the clouds possess a random, internal field line structure (which they
might do if they were pre--existing entities that were swept up by the
jet rather than being formed from condensations within the jet), then the
tangential component of the internal field lines can prevent the
infiltration of fast particles from outside the clouds.

It has been pointed out, however, that collective plasma effects, triggered
by the passage of fast particles through a `cold' plasma, can enhance the
heating rate, eqn. (\ref{eq:Coulheat}), by a factor as large as $10^5$
(see \cite{FerlMush84} and references therein).
If this is the case, then the only way cool clouds can maintain their
properties in the presence of fast particles is to efficiently radiate
away any extra energy input.
The dominant radiative cooling process in a typical BAL cloud is through
the {\small CIV}~$\lambda 1549$ line transition, which has an Einstein
coefficient $A_{21} \simeq (2.6 \times 10^7 ) \, {\rm s}^{-1}$.
The volume cooling rate for this line transition is then $\nciv A_{21}
h\nu_{21} \simeq (3 \times 10^{-4}) \chiciv n_{\rm cld}$~erg.s
$^{-1}$cm$^{-3}$, where $\chiciv = \nciv / n_{\rm cld}$ is the abundance
of the CIV ion relative to the total (ionized plus neutral) hydrogen density
in the clouds.
The total heating can be quantitatively estimated as
$\simeq \zeta H^{\rm Coul}$, where $\zeta \ltapprox 10^5$ takes into account
collective plasma heating and where $H^{\rm Coul}$ is integrated over the
nonthermal electron distribution (neglecting the weak $\gamma$--dependence
in the $\cal B$ parameter).
The ratio of the cooling to heating rates is then
\begin{equation}
\frac{C^{\scriptscriptstyle {\rm CIV}}}{H^{\rm tot}} \sim 10^6 \,
n_{\rm jet}^{-1} \, \zeta_5^{-1} \left( \frac{\chiciv}{10^{-4}} \right)
\;\; {\rm cm}^{-3} \;\; ,
\end{equation}
where $\zeta_5 = \zeta / 10^5$ and where an appropriate value of
${\cal B} = 25$ has been used.
Thus, radiative line cooling in the clouds is efficient enough to overcome
any extra heat input from the ambient relativistic jet plasma provided its
density is $n_{\rm jet} \ll 10^6$~cm$^{-3}$.
According to the limits on $n_{\rm jet}$ imposed by the total jet energy
budget, eqn. (\ref{eq:njet}), this is always satisfied and therefore,
evaporation does not pose an immediate threat to the survival of BAL clouds
embedded within a weak, sub--relativistic jet.

\section{Summary and Discussion}
\label{sec:conc}

It has been argued that the phenomenon of nuclear absorption outflows
from quasars provides an important clue to understanding the observed
radio--loud/radio--quiet dichotomy in active galactic nuclei if
interpreted in terms of an inhomogeneous weak jet model in which
the thermal gas responsible for the observed UV absorption troughs is
embedded within a poorly--collimated outflow of weakly radio--emitting
plasma.
The motivation for this model is threefold: (i) observations of radio--quiet
quasars have confirmed that the nature of their radio emission is
fundamentally similar to that of radio--loud quasars; (ii) the observed
anticorrelation between the terminal velocity of outflowing thermal gas
and the radio strength of the quasar is direct evidence that the dynamics
of the thermal, absorbing gas is intimately linked to the properties of
the nonthermal, radio--emitting plasma; and (iii) lower velocity intrinsic
absorption outflows have also been detected in Seyferts, many of which
are found to exhibit linear radio structure indicative of small--scale,
weak jets.

The observational constraints obtained here corroborate other theoretical
jet models which attribute the differences in radio strength (i.e. quiet,
weak and loud) to differences in the physical properties of jets (e.g. total
energy flux, bulk speed, relative quantities of thermal and nonthermal plasma).
It has also been suggested that a weak jet interpretation of the observed
radio flux may offer a viable explanation for why nuclear absorption outflows
are not detected in strong
radio sources; the relativistic jets which power these sources are thought to
be propitious sites for {\it in situ} particle acceleration, which precludes
the accumulation of cooled particles that can thermalize and condense to form
localized gas clouds capable of producing absorption features.
Furthermore, since the efficiency of ram--pressure acceleration increases as
a jet becomes sub--relativistic, the observed anticorrelation between the
terminal velocity of the outflowing thermal gas and the radio strength of
the quasar can also be explained by a weak jet model.
Finally, it has been demonstrated that this model can explain the narrow
UV absorption features detected in some Seyfert~1s.
Thus, it has been shown that a weak jet model is successful in explaining not
only the high velocity outflows in bonafide broad absorption line quasars,
but also the lower velocity nuclear absorption outflows detected in both
their strong radio counterparts and low luminosity counterparts.

The importance of broad absorption line quasars to our understanding of
the radio--loud/radio--quiet dichotomy becomes evident in the framework
of a weak jet model, of which the underpinning implication is that all
AGN possess radio--emitting jets to varying degrees and that their
observational classification depends not only upon orientation, but also
upon the intrinsic differences in the physical properties of their jets.
However, one key issue which is yet to be fully resolved is the role
evolutionary effects play in the formation of jets and outflows in AGN;
if mass ejection phenomena are evolutionary phases, then the true
significance of broad absorption line quasars in the grand scheme
of AGN unification is yet to be fully appreciated.

\acknowledgments{The author wishes to thank the Royal Commission for the
Exhibition of 1851 Research Fellowship (Imperial College, London) for
financial support and A.~C. Gower and G.~F. Lewis for helpful discussions.}

\end{document}